\documentclass[aps,showpacs,11pt,superscriptaddress,preprintnumbers,amsmath,amssymb,nofootinbib]{revtex4-1}
\pdfoutput=1
\usepackage{mathrsfs}
\usepackage{epsfig}
\usepackage{graphicx}
\usepackage{dcolumn}
\usepackage{bm}
\usepackage{amsmath}
\usepackage{slashed}       
\usepackage{amssymb}
\usepackage{amsfonts}

\usepackage[T1]{fontenc}
\usepackage{multirow}
\usepackage{booktabs}
\usepackage[usenames,dvipsnames,svgnames,table]{xcolor}
\usepackage{siunitx} 
\usepackage{url} 

\definecolor{interesting}{rgb}{0., 1., 0.5}
\definecolor{suppressed}{rgb}{0.97, 0.51, 0.47}

\newcommand{\gev}{\ensuremath{\,\text{GeV}}}

\newcommand{\lsp}{{\tilde{\chi}_1^0}}

\begin{document}

\preprint{CoEPP-MN-18-5}
\title{Constraining resonant dark matter with combined LHC electroweakino searches}
\author{Giancarlo Pozzo}
\email{giancarlo.pozzo@monash.edu}
\affiliation{ARC Centre of Excellence for Particle Physics at the Tera-scale, School of Physics and Astronomy, Monash University, Melbourne, Victoria 3800, Australia}
\author{Yang Zhang}
\email{yang.zhang@monash.edu}
\affiliation{ARC Centre of Excellence for Particle Physics at the Tera-scale, School of Physics and Astronomy, Monash University, Melbourne, Victoria 3800, Australia}

\begin{abstract}
    In the Minimal Supersymmetric Standard Model light neutralinos can satisfy the dark matter (DM) abundance constraint by resonant annihilation via a $Z$ or a light Higgs ($h$) boson.
    In this work we study the current and future status of this scenario by investigating relevant experimental constraints, including DM direct detection, measurements of $Z$ and Higgs invisible decays, and direct searches at the Large Hadron Collider (LHC).
    To take full advantage of the LHC data, we combine the results of all relevant electroweakino searches performed by the Compact Muon Solenoid (CMS) Collaboration.
    Such combination increases the bound on the Higgsino mass parameter to $|\mu|>390$ GeV, which is about 80 GeV stricter than the bound obtained from individual analyses.
    In a simplified model we find that the $Z$ funnel region is on the brink of exclusion, the $h$ funnel for $\mu<0$ only survives if $\tan\beta<7.4$, and the $h$ funnel for $\mu>0$ is the main surviving region.
    Future DM direct detection experiments, such as LUX and ZEPLIN, can explore the whole region, while the high luminosity LHC can exclude $\tan\beta>8$ for $\mu>0$ and $\tan\beta>5.5$ for $\mu<0$.
    After applying the muon anomalous magnetic moment constraint only a tiny part of the $Z/h$ funnel region survives which will soon be probed by ongoing experiments.
\end{abstract}

\maketitle

\section{Introduction}\label{sec:intro}

A wide range of astrophysical observations indicates the existence of dark matter (DM) at various length scales via gravitational effects.  Motivated by this during the last decades considerable effort was made to detect DM particles at collider experiments (such as LEP~\cite{Abdallah:2003np} and the LHC~\cite{Aaboud:2017phn,Sirunyan:2017hci}), in direct (by XENON1T~\cite{Aprile:2017aty}, LUX~\cite{Akerib:2015wdi} or PandaX~\cite{Cui:2017nnn}) and indirect (AMS-II~\cite{Aguilar:2013qda}, Fermi-LAT~\cite{Ackermann:2015zua} or DAMPE~\cite{TheDAMPE:2017dtc}) detection experiments.  Despite the lack of direct experimental evidence, the lightest neutralino of the R-parity conserving Minimal Supersymmetric Standard Model (MSSM)~\cite{Athron:2017yua, Bagnaschi:2017tru, Aad:2015baa} remains an especially attractive DM candidate.  This is because, beyond dark matter, the MSSM provides solutions to several problems of the Standard Model (SM): the lightness of the observed Higgs mass, a dynamical mechanism of electroweak symmetry breaking, the unification of particles and forces and beyond.

Supersymmetric (SUSY) global fits, which also include experimental constraints on DM particles, have delineated the most likely model parameter regions~\cite{Athron:2017qdc, Athron:2017yhy, Athron:2017ard, Ajaib:2017iyl, Han:2016gvr, Bechtle:2015nua, Bagnaschi:2015eha, Balazs:2013qva, Fowlie:2012im, Buchmueller:2012hv, Strege:2012bt, Citron:2012fg, Bornhauser:2013aya, Henrot-Versille:2013yma, Bechtle:2013mda, Buchmueller:2013rsa, Ellis:2013oxa, Bechtle:2014yna, Bechtle:2012zk, Bagnaschi:2017tru, Athron:2017yua, Aad:2015baa}.  In global fits of the phenomenological MSSM, there is always a $Z/h$ funnel region in which neutralino dark matter can achieve the right thermal relic density through $Z$ or Higgs boson resonant annihilation.  Consequently, in this region the DM mass is about half of the $Z$ or Higgs boson mass.  Comparing to other regions, the $Z/h$ funnel region is an islet in the parameter space where some of the supersymmetric particles (sparticles) are relatively light. These characteristics make the sparticles in the $Z/h$ funnel region the most promising candidates to be detected at the LHC and DM search experiments.
More importantly, several modest excesses of data above the expected background were found in the signal regions of recent CMS and ATLAS electroweakino searches, including signal region \texttt{SR3$\ell$\_ISR} (3.02 $\sigma$ deviation), \texttt{SR3$\ell$\_LOW} (2.13 $\sigma$ deviation) and \texttt{SR2$\ell$\_ISR} (1.99 $\sigma$ deviation) in ATLAS recursive jigsaw reconstruction analysis~\cite{Aaboud:2018sua},  \texttt{SR0D} (2.3 $\sigma$ deviation) in ATLAS $\geq 4 \ell+E_T^{miss}$ analysis~\cite{Aaboud:2018zeb}, and the not-$tt$-like signal region for masses between 96 and 150 GeV (2.0$\sigma$ deviation) in CMS $2 \ell+E_T^{miss}$ analysis~\cite{Sirunyan:2017qaj}. The global fit of the electroweakino sector performed by GAMBIT Collaboration shows that the $Z/h$ funnel region is  consistent with a new physics interpretation of these excesses~\cite{Kvellestad:2018akf,Athron:2018vxy}. Motivated by these results, in this work we carefully explore the present and future status of the $Z/h$ funnel region.

On the theoretical side, $Z/h$ resonant annihilation is important in natural SUSY~\cite{Baer:2012up}, especially in the natural MSSM, since it allows the lightest neutralino to achieve the observed thermal relic density~\cite{Abdughani:2017dqs}.  In the natural Next-to-MSSM (NMSSM), although the inclusion of a singlet superfield relaxes the experimental constraints on the electroweakinos, the exclusion of the $Z/h$ funnel region increases the lower limit on the DM mass from \SI{20}{\GeV} to \SI{80}{ \GeV}~\cite{Cao:2016nix, Cao:2016cnv, Ellwanger:2016sur}.
The lower limit on the DM mass, in turn, is critical for any LHC sparticle search because under R-parity all sparticles decay to the lightest supersymmetric particle (LSP) $\lsp$ and the LSP mass is folded into the analyses.  Typically, stricter search limits arise
in analyses
with light
neutralinos.  In a simplified model, for instance, with first- and second-generation mass-degenerate squarks, squark masses below \SI{1.6}{\TeV} (\SI{1.4}{\TeV}) are excluded for $m_\lsp<\SI{200}{\GeV}$ ($ \SI{200}{\GeV} < m_\lsp<\SI{400}{\GeV}$), but entirely survive if $m_\lsp>\SI{600}{\GeV}$~\cite{Aaboud:2017vwy}.  Therefore, in most cases, the exclusion of the $Z/h$ funnel region affects the mass limits of all sparticles.

The MSSM $Z/h$ funnel region have been examined in numerous recent papers~\cite{Han:2016qtc, Calibbi:2014lga, vanBeekveld:2016hug, Caron:2015wda, Barman:2017swy, Bramante:2015una, Chakraborti:2017vxz, Badziak:2017uto, Han:2013kza, Badziak:2015exr, Choudhury:2017acn, Chakraborti:2017dpu, Kobakhidze:2016mdx, Choudhury:2016lku, vanBeekveld:2016hbo, Profumo:2017ntc, Calibbi:2011ug, Belanger:2001am, Han:2013gba, Xiang:2016ndq, Hamaguchi:2015rxa, Han:2014sya, new, Arina:2016rbb, Martin:2014qra, Barducci:2015ffa}. 
The constraints from LHC Run-I SUSY direct searchers were implemented by requiring that the SUSY signal events do not exceed the $95\%$ confidence level (C.L.) upper limit in the signal region with the best-expected exclusion power~\cite{Hamaguchi:2015rxa, Han:2014sya, Calibbi:2014lga, Barman:2017swy}.
At \mbox{Run-I}, due to relatively small backgrounds of leptonic processes, the signal region with the best-expected exclusion power for the $Z/h$ funnel region comes from the "3$\ell$" search for the $pp\to \tilde{\chi}_{1}^\pm\tilde{\chi}_{2}^0\to W^{\pm}Z\tilde{\chi}_{1}^0\tilde{\chi}_{1}^0 \to \ell\ell v\ell \tilde{\chi}_{1}^0\tilde{\chi}_{1}^0$ process~\cite{Aad:2014nua}.
However, with the increase of centre-of-mass-energy and integrated luminosity, the boosted jets can also be used to distinguish signals of heavy electroweakinos from background events.
As a result, the sensitivities of searches for other decay modes will increase significantly, even surpassing the "3$\ell$" search.
An example is the "1$\ell2b$" search for the $pp\to \tilde{\chi}_{1}^\pm\tilde{\chi}_{2}^0\to W^{\pm}H\tilde{\chi}_{1}^0\tilde{\chi}_{1}^0 \to b\bar{b}v\ell \tilde{\chi}_{1}^0\tilde{\chi}_{1}^0$ process with one lepton, two b-jets and $E_T^{\rm miss}$ final state.
At the high luminosity LHC (HL-LHC), the $95\%$ C.L. exclusion contour of "3$\ell$" search reaches \SI{1100}{\GeV} in the case of the $WZ$-mediated simplified models~\cite{ATL-PHYS-PUB-2014-010}, while the exclusion contour of "1$\ell2b$" search reaches \SI{1310}{\GeV} in $\tilde{\chi}_{1}^\pm$, $\tilde{\chi}_{2}^0$ mass in the case of the $Wh$-mediated simplified models using the MVA technique~\cite{ATL-PHYS-PUB-2015-032}.
At Run-II the impact of "1$\ell2b$" search in the $Z/h$ funnel region cannot be ignored, because $\tilde{\chi}_{1}^\pm$ decay exclusively to $\tilde{\chi}_{1}^0 W^{\pm}$ while  BR$(\tilde{\chi}_{2}^0\to\tilde{\chi}_{1}^0 h)+$BR$(\tilde{\chi}_{3}^0\to\tilde{\chi}_{1}^0 h)\simeq 90\%$~\cite{Calibbi:2014lga}.
A statistical combination of exclusive signal regions in these searches maximizes the discovery potential.
For example, in the case of the $WZ$-mediated simplified models, the combination performed by CMS~\cite{Sirunyan:2018ubx} improves on the "$3\ell$" analysis yielding an observed lower limit of \SI{150}{\GeV} on the chargino mass.

In this work, we study the present status of $Z/h$ funnel region under the constraint of $3l+E_T^{miss}$~\cite{Sirunyan:2017lae}, $2l+E_T^{miss}$~\cite{Sirunyan:2017qaj} and $1l+2b+E_T^{miss}$~\cite{Sirunyan:2017zss} searches using \SI{13}{\TeV}~\SI{35.9}{\femto\barn^{-1}} LHC data, as well as the latest DM direct detection results.  The rest of paper is organized as follows.  In Section~\ref{sec:Models} we briefly describe the electroweakino sector of MSSM, with focus on the properties of DM.  We present the parameter space of the $Z/h$ funnel region and related constraints in Section~\ref{sec:constraints}.  The HL-LHC reach for the regions that survive the present LHC constraints is discussed in Section~\ref{sec:HL-LHC}. In Section~\ref{sec:pmssm} we investigate the $Z/h$ funnel region in a practical phenomenology model.  Finally, we draw our conclusions in Section~\ref{sec:summary}.


\section{The $Z/h$-resonant neutralino dark mater}
\label{sec:Models}

In this section we describe the MSSM electroweakino sector, that is the superpartners of the electroweak gauge bosons (Bino $\tilde{B}$ and Winos $\tilde{W}$) and the two Higgs doublets (Higgsinos $\tilde{H}$).
After electroweak symmetry breaking the electroweakinos mix to form neutralino $\tilde{\chi}_{i}^0~ (i=1,2,3,4)$ and chargino $\tilde{\chi}_{i}^\pm~(i=1,2)$ mass eigenstates.
In the $\psi_{\alpha}=(\tilde{B},\tilde{W}^0,\tilde{H}_d^0,\tilde{H}_u^0)$ basis neutralino masses are given by $-\frac{1}{2}[\psi_{\alpha}{\cal{M}_{\widetilde{\chi}^0}}_{\alpha\beta}\psi_{\beta}+h.c.]$ with the non-diagonal, symmetric mass matrix
\begin{equation}
{\cal{M}}_{\widetilde{\chi}^0} = \left( \begin{array}{cccc}
  M_1 			      & 0 				      & -M_Z {\rm s}_W {\rm c}_\beta  &  M_Z {\rm s}_W {\rm s}_\beta \\
  0				      & M_2 			      &  M_Z {\rm c}_W {\rm c}_\beta  & -M_Z {\rm c}_W {\rm s}_\beta \\
  -M_Z {\rm s}_W {\rm c}_\beta &  M_Z {\rm c}_W {\rm c}_\beta  & 0 				      & -\mu 			      \\
   M_Z {\rm s}_W {\rm s}_\beta & -M_Z {\rm c}_W {\rm s}_\beta &  -\mu 			      & 0				      \\
   \end{array} \right) .
\end{equation}
Here $M_1$, $M_2$ and $\mu$ are the Bino, Wino and Higgsino masses, ${\rm s}_\beta = \sin\beta$ and ${\rm c}_\beta = \cos\beta$ where $\tan\beta = \langle H_u\rangle/\langle H_d\rangle$ is the ratio of the vacuum expectation values of the two Higgs doublets, $M_Z$ is the $Z$ boson mass, and ${\rm s}_W$ and ${\rm c}_W$ are the sine and cosine of the weak mixing angle $\theta_W$.  With the same notation, in the $(\tilde{W}^{\pm},\tilde{H}^{\pm})$ basis the chargino mass matrix is given by
\begin{equation}
{\cal{M}}_{\widetilde{\chi}^\pm} = \left( \begin{array}{cc}
M_2  & \sqrt{2}{\rm c}_\beta M_W \\
\sqrt{2}{\rm s}_\beta M_W &  \mu \\
\end{array} \right) ,
\end{equation}
where $M_W$ is the W boson mass.  The physical masses of the neutralinos and charginos are given by the eigenvalues of ${\cal{M}}_{\widetilde{\chi}^0}$ and ${\cal{M}}_{\widetilde{\chi}^\pm}$.

Due to the $m_{\tilde{\chi}_1^{\pm}} > \SI{92}{\GeV}$ chargino mass limit from LEP~\cite{LEP}, the Wino mass, $M_2$, and Higgsino mass, $|\mu|$, must be higher than about $\SI{100}{\GeV}$.  As a result, the lightest neutralino, with mass $m_{\tilde{\chi}_1^0}\sim M_Z/2$ or $M_{h}/2$, must be Bino dominated.  We demand it to be the LSP, and R-parity conservation renders it a DM candidate.
The main annihilation mode for this DM proceeds via an $s$-channel $Z$ or Higgs boson, and the corresponding annihilation cross section is given by~\cite{Hamaguchi:2015rxa}:
\begin{equation} \label{annihilation}
\sigma(\tilde{\chi}_1^0\tilde{\chi}_1^0\to Z/h\to f \bar{f}) \simeq
\frac{1}{2} C_{Z/h}^2 \sqrt{1-\frac{4m_{\tilde{\chi}_1^0}^2}{s}} \frac{1}{(s-M_{Z/h}^2)^2+(M_{Z/h}\Gamma_{Z/h})^2} \frac{s}{M_{Z/h}} \Gamma_{Z/h\to f\bar{f}} ,
\end{equation}
where $C_{Z/h}$ is the coupling between $\tilde{\chi}_1^0$ and the $Z/h$ boson, and $\Gamma_{Z/h}$ is the corresponding decay width.  The couplings arise via neutralino mixing, as shown by the relevant Lagrangian term~\cite{Cao:2015efs}:
\begin{eqnarray} \label{interaction}
{\cal{L}}_{\tilde{\chi}^0} & = &  \frac{e}{ {\rm s}_W} h \bar{\tilde{\chi}}_1^0
 (N_{12} - N_{11} \tan \theta_W ) (\sin \alpha N_{13} + \cos \alpha N_{14} ) {\tilde{\chi}}_1^0  \nonumber \\
&& + \frac{e}{{{\rm s}_W {\rm c}_W}} Z_\mu \bar{\tilde{\chi}}_i^0 \gamma^\mu\Big[ \frac{P_L}{2}(N_{14}^2-N_{13}^2) + \frac{P_R}{2}(N_{14}^2-N_{13}^2) \Big] {\tilde{\chi}}_j^0 .
\end{eqnarray}
Here $\alpha$ is the Higgs mixing angle, and $N_{ij}$ are the elements of the $4 \times 4$ unitary matrix that diagonalizes the neutralino mass matrix ${\cal{M}}_{\widetilde{\chi}^0}$ such that $N_{11}^2$, $N_{12}^2$ and $N_{13,14}^2$ are the Bino, Wino and Higgsino components of $\tilde{\chi}_1^0$, respectively.
Equation~\eqref{interaction} shows that the Higgsino components play an important role both in the $h\tilde{\chi}_1^0\tilde{\chi}_1^0$ and $Z\tilde{\chi}_1^0\tilde{\chi}_1^0$ interaction.

Considering the limit $M_1<\SI{100}{\GeV}< |\mu|\ll M_2$, the Higgsino components can be expressed as~\cite{Calibbi:2014lga}
\begin{eqnarray} \label{component}
N_{13}=\frac{M_Z{\rm s}_W}{\mu}\left({\rm s}_{\beta}+{\rm c}_{\beta}\frac{M_1}{\mu}\right),\qquad
N_{14}=-\frac{M_Z{\rm s}_W}{\mu}\left({\rm c}_{\beta}+{\rm s}_{\beta}\frac{M_1}{\mu}\right),
\end{eqnarray}
which decrease when the mass hierarchy between Higgsino and Bino increases.  From equations~\eqref{component} and~\eqref{interaction}, one can derive the couplings
\begin{eqnarray} \label{coupling}
C_Z=\frac{e M_Z^2}{\mu^2}\cos(2\beta) \left( 1+\frac{M_1^2}{\mu^2} \right),\qquad
C_h=\frac{e M_Z}{\mu}\left[ \cos(\beta+\alpha) +\sin(\beta-\alpha)\frac{M_1}{\mu} \right].
\end{eqnarray}
Thus, the relic density of $Z/h$-resonant DM at tree level depends on $M_1$, $\mu$ and $\tan\beta$.  We, therefore, perform a scan over $M_1$, $\mu$ and $\tan\beta$ to identify the parameter space where $Z/h$-resonant DM satisfies the observed DM abundance.  Following that, we examine the impact of current and future experimental constraints on this parameter space.

\section{The parameter space and constraints}
\label{sec:constraints}
To analyse the $Z/h$ funnel region, we first study a simplified model that assumes the sfermion masses, wino mass $M_2$, gluino mass $M_3$ and CP-odd Higgs mass $M_A$ are fixed at \SI{3}{\TeV}, heavy enough to decouple at LEP or the LHC. We set all the trilinear couplings except $A_t$ to zero.
To match the measured value of SM-like Higgs mass of \SI{125.09}{\GeV}~\cite{Aad:2015zhl}, the trilinear coupling $A_t$ is fixed at \SI{4.5}{\TeV} for $\tan \beta>10$, at \SI{5.0}{\TeV} for $7< \tan \beta<10$ and at \SI{6.0}{\TeV} for $\tan \beta<7$.
Under these assumptions, we sample the following parameter space:
\begin{eqnarray}
\label{eq:scanRange}
10~{\rm GeV}<M_1<100~{\rm GeV}, \qquad 50~{\rm GeV}<|\mu|<1500~{\rm GeV}, \qquad 5<\tan \beta<50 . 
\end{eqnarray}
We use \texttt{SUSY-HIT-1.5}~\cite{Djouadi:2006bz} based on \texttt{SuSpect}~\cite{Djouadi:2002ze}, together with \texttt{SDECAY} \cite{Djouadi:2006bz,Muhlleitner:2004mka} and \texttt{HDECAY} \cite{Djouadi:1997yw} to generate the mass spectrum and to calculate the $Z/h$ boson decay branching ratios, \texttt{micrOMEGAs-4.3.5} \cite{Belanger:2001fz,Barducci:2016pcb} to calculate the DM observables, and \texttt{EasyScan\_HEP}~\cite{Han:2016gvr} to perform the scan.  Due to the low dimensionality and simplicity of the parameter space we generate samples on a grid.

In Sections \ref{sec:relic}-\ref{sec:13lhc} we detail the relevant constraints on the $Z/h$-resonant DM.  Here we ignore other observations, such as B-physics measurements, that tend to give mild constraints due to the high scale of the fixed SUSY parameters.

\begin{figure}[th]
\centering 
\includegraphics[width=.9\textwidth]{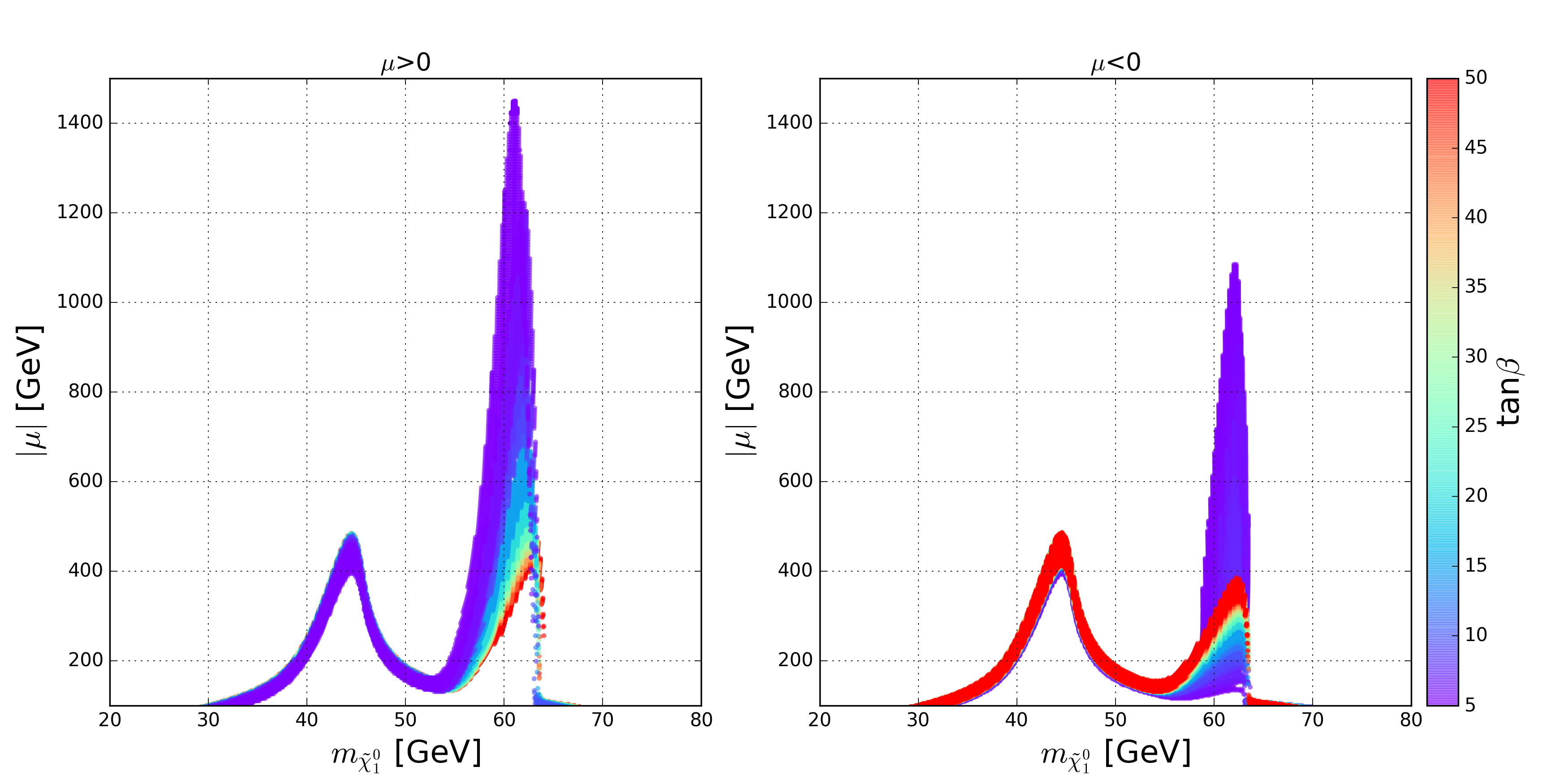}
\caption{\label{fig:mu_0} Parameter regions allowed by the observed DM relic density ($0.0959<\Omega h^2<0.1439$) on the Higgsino mass vs. lightest neutralino mass plane for $\mu>0$ (left panel) and $\mu<0$ (right panel).  Colours show the value of $\tan\beta$.  The masses of sparticles other than the electroweakinos are fixed at \SI{3}{\TeV}.  The value of $A_t$ is also fixed to obtain the observed Higgs mass: $A_t=\SI{4.5}{\TeV}$ for $\tan\beta>10$, $A_t=\SI{5.0}{\TeV}$ for $7<\tan\beta<10$ and $A_t=\SI{6.0}{\TeV}$ for $\tan\beta<7$. }
\end{figure}

\subsection{The thermal relic density of DM}\label{sec:relic}

From equations~\eqref{coupling} and~\eqref{annihilation}, we see that the measurement of the DM abundance by Planck~\cite{Ade:2015xua} and WMAP~\cite{Dunkley:2008ie} place severe restrictions on the relationship among $M_1$, $\mu$ and $\tan\beta$.  We assume that the thermal relic density of the lightest neutralino is equal to the cold DM abundance $\Omega h^2=0.1199 \pm 0.0022$ at $2\sigma$ level with 10\% theoretical uncertainty (c.f. the \texttt{Plik} cross-half-mission likelihood in~\cite{Ade:2015xua}).  In Figure~\ref{fig:mu_0} we project the allowed regions on the $(m_{\tilde{\chi}_1^0},|\mu|)$ plane for both $\mu>0$ and $\mu<0$ with colours indicating the value of $\tan\beta$.

As sketched in Section~\ref{sec:Models}, to achieve both the observed DM abundance and a sizeable coupling to the $Z/h$ boson, the Bino-like $\tilde{\chi}_1^0$ must contain a
certain amount of
Higgsino component.  This imposes limits on the Higgsino mass, shown in Figure~\ref{fig:mu_0} by the coloured regions.  The blank region above the coloured region leads to an overproduction of DM in the early universe, while the blank region below the coloured region has a relic density smaller than $0.096$.  Due to the resonance in equation~\eqref{annihilation}, the Higgsino mass is enhanced when $m_{\tilde{\chi}_1^0}$ close to $M_{Z/h}$, therefore the allowed region features two clear peaks.

The Higgs resonances (the peaks around $m_h/2$) in the left ($\mu>0$) and right ($\mu<0$) panel of Figure~\ref{fig:mu_0} show different dependence on $\tan\beta$ for a fixed $m_\lsp$.  This difference is caused by the sign of $M_1/\mu$ in the coupling between the $\tilde{\chi}_1^0$ and the Higgs boson.  Taking the decoupling limit of the Higgs sector, $\beta-\alpha=\pi/2$, $C_h$ in equation~\eqref{coupling} can be written as
\begin{eqnarray} \label{coupling_h}
C_h=\frac{e M_Z}{\mu} \left( \sin2\beta +\frac{M_1}{\mu} \right).
\end{eqnarray}
Therefore, for $M_1/\mu>0$ and $M_1\simeq M_h/2$ to keep the coupling $C_h$ unchanged the Higgsino mass has to increase from \SI{400}{\GeV} to \SI{1440}{\GeV} and $\tan\beta$ has to decrease from $50$ to $5$.  For the same reason, for $M_1/\mu<0$ and $M_1\simeq M_h/2$ the coupling is bracketed as $|\mu|$ decreases from \SI{380}{\GeV} to \SI{130}{\GeV} and $\tan\beta$ decreases from 50 to 7.  For $M_1/\mu<0$ and $\tan\beta<7$ there are two separate regions corresponding to the observed relic density, divided by the so-called "blind spot" where $\sin 2\beta=M_1/\mu$~\cite{Cheung:2012qy, Huang:2014xua, Badziak:2015exr, Badziak:2016qwg, Cao:2016cnv}.  The coupling $C_h$ changes sign between the two regions.  For $\tan\beta=5$ and $m_{\tilde{\chi}_1^0}=\SI{52}{\GeV}$, for example, the regions $\mu<\SI{-136}{\GeV}$ and $\SI{-168}{\GeV} <\mu< \SI{-1085}{\GeV}$ both correspond to $\Omega h^2<0.14$.

The $Z$ resonance, on the other hand, is independent of the sign of $M_1/\mu$ and it mildly depends on $\tan\beta$, as shown in equation~\eqref{coupling}.  The Higgsino can be as heavy as about \SI{470}{\GeV} when DM annihilates via the $Z$ resonance.

\subsection{Dark matter direct detection experiments}
\label{sec:DMdetection}

\begin{figure}[th]
\centering 
\includegraphics[width=.9\textwidth]{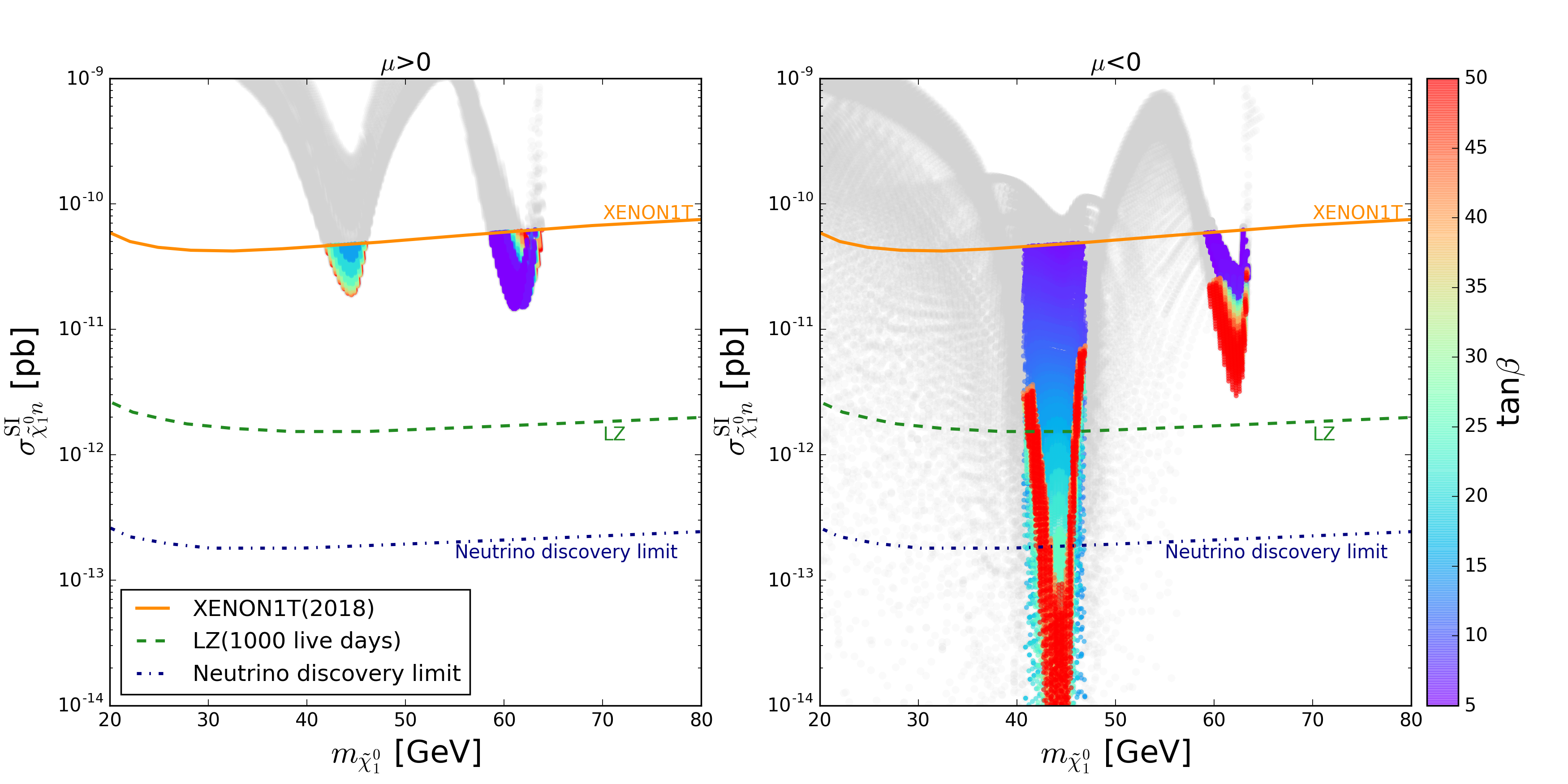}\\
\includegraphics[width=.9\textwidth]{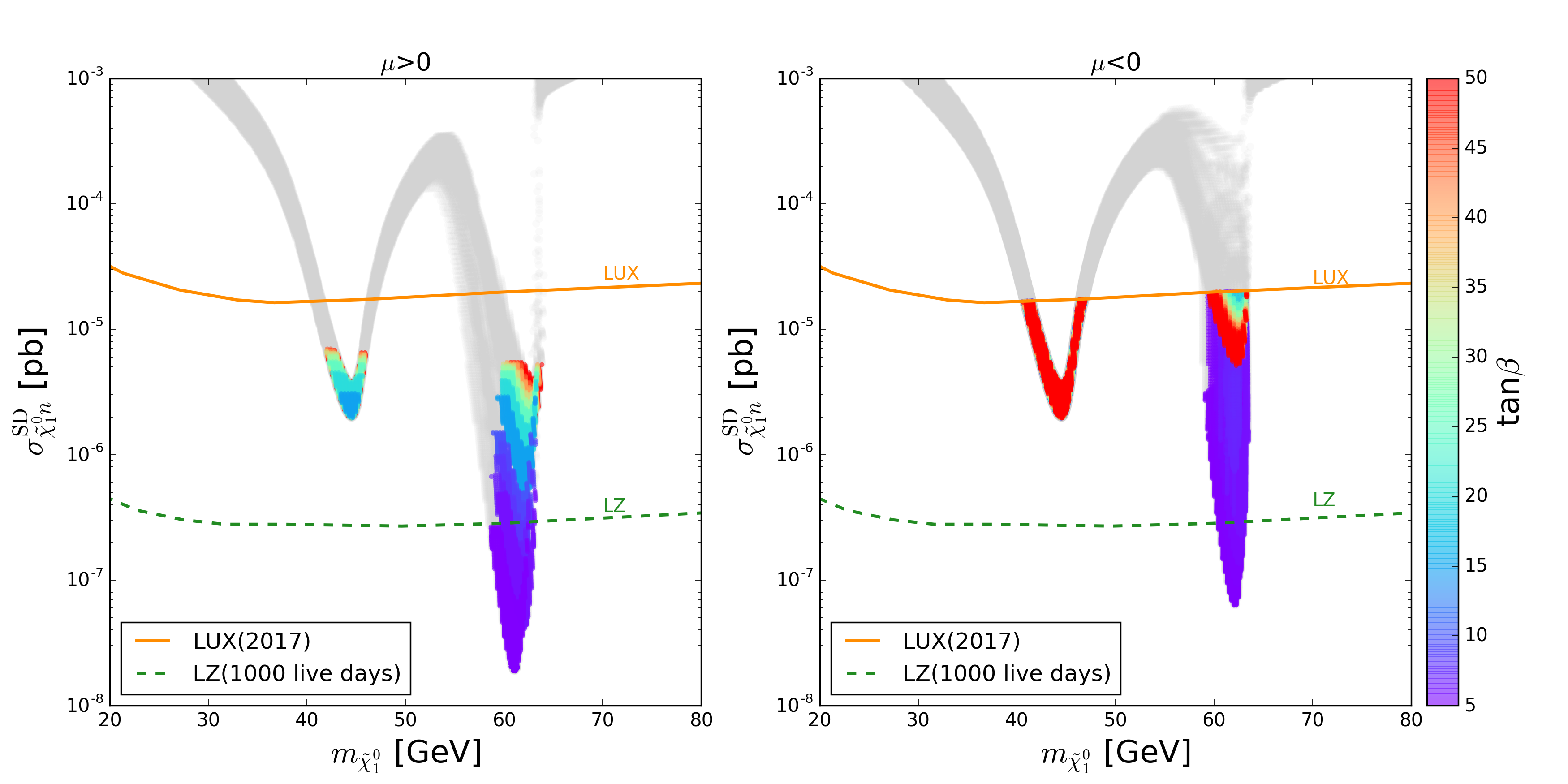}
\caption{\label{fig:DD} Parameter regions allowed by the observed DM abundance ($0.0959<\Omega h^2<0.1439$) on the $(m_{\tilde{\chi}_1^0},\sigma_{\tilde{\chi}_1^0n}^{\rm SI})$ logarithmic plane (upper panels) and $(m_{\tilde{\chi}_1^0},\sigma_{\tilde{\chi}_1^0n}^{\rm SD})$ logarithmic plane (lower panels) for $\mu>0$ (left panels) and $\mu<0$ (right panels).
The orange solid lines mark the limit on $\sigma_{\tilde{\chi}_1^0n}$ given by XENON1T~\cite{Aprile:2017iyp,xenon1t_2018} and LUX~\cite{Akerib:2017kat} experiments. The green dashed lines mark the projected limit of LUX-ZEPLIN~\cite{Akerib:2018lyp}.
The colours show the value of $\tan\beta$; grey regions are excluded by DM direct detection at 90\% C.L. }
\end{figure}

Neutralinos with non-negligible Higgsino component can be directly detected via elastic scattering on nuclei mediated by $Z$ or Higgs boson exchange~\cite{Akerib:2016vxi, Aprile:2017iyp, Cui:2017nnn, Fu:2016ega, Aprile:2016swn, Akerib:2017kat}.  The null result of the searches for such scattering by LUX~\cite{Akerib:2016vxi}, XENON1T~\cite{Aprile:2017iyp,xenon1t_2018} and PandaX-II~\cite{Cui:2017nnn} provides limits on the spin-independent (SI) neutralino-nucleon elastic cross section $\sigma_{\tilde{\chi}_1^0n}^{\rm SI}$.  In the $\tilde{\chi}_1^0$ mass region we consider the one-sided 90\% C.L. upper limit on $\sigma_{\tilde{\chi}_1^0n}^{\rm SI}$ is about \SI{5e-11}{\pico\barn} ~\cite{xenon1t_2018}.  The most sensitive constraints on spin-dependent (SD) DM-neutron elastic cross section $\sigma_{\tilde{\chi}_1^0n}^{\rm SD}$ and DM-proton elastic cross section $\sigma_{\tilde{\chi}_1^0p}^{\rm SD}$ come from LUX~\cite{Akerib:2017kat} and PICO-60~\cite{Amole:2017dex}, respectively.  In Figure~\ref{fig:DD} we show current, as well as projected LUX-ZEPLIN(LZ)~\cite{Akerib:2018lyp}, constraints on $\sigma_{\tilde{\chi}_1^0n}^{\rm SI}$ and $\sigma_{\tilde{\chi}_1^0n}^{\rm SD}$
in the parameter regions that account for the observed DM abundance.  The grey regions are excluded by either DM SI or SD scattering searches.

The top panels of Figure~\ref{fig:DD} show the predicted $\sigma_{\tilde{\chi}_1^0n}^{\rm SI}$ in the surviving region as a function of $m_{\tilde{\chi}_1^0}$.  In the limit of heavy scalar superpartners, the dominant contribution of  $\sigma_{\tilde{\chi}_1^0n}^{\rm SI}$ comes from the t-channel exchange of a Higgs boson~\cite{Badziak:2015exr, Badziak:2016qwg}:
\begin{equation}
\sigma_{\tilde{\chi}_1^0n}^{\rm SI} \simeq\frac{4\mu_r^2}{\pi}
\Big[\sum^{2}_{i=1}\frac{C_{h_i\tilde{\chi}_1^0\tilde{\chi}_1^0} C_{h_iNN} }{2M_{h_i}^2} \Big]^2 .
\end{equation}
Here $\mu_r$ is the neutralino-nucleus reduced mass, $C_{h_iNN}$ denotes the effective coupling between the  Higgs and nucleon.
As discussed in Subsection~\ref{sec:relic}, in the vicinity of the Higgs resonance $C_{h\tilde{\chi}_1^0\tilde{\chi}_1^0}$ is restricted by the observed DM abundance.
In this region $\sigma_{\tilde{\chi}_1^0n}^{\rm SI}$ is practically independent of $\tan\beta$ and sign of $\mu$, and it is large enough to be fully covered by the LZ projected limits.  On the other hand, on the $Z$ resonance the DM relic density is independent of $C_{h\tilde{\chi}_1^0\tilde{\chi}_1^0}$, and demands a fixed $|\mu|$ for certain $m_{\tilde{\chi}_1^0}$, such as $|\mu|\simeq\SI{450}{\GeV}$ for $m_{\tilde{\chi}_1^0}=\SI{45}{\GeV}$.  As a result, for $\mu>0$ the $\sigma_{\tilde{\chi}_1^0n}^{\rm SI}$ cross section decreases when $\tan\beta$ increases and will be detectable at LZ.  For $\mu<0$, however, due to the blind spot at $\sin2\beta=M_1/\mu$, it is impossible to test $Z$-resonance DM for $\tan\beta={\rm tan[arcsin(45/450)/2]}\simeq20$.

On the contrary, at tree level and in the heavy squark limit only the t-channel $Z$ boson exchange diagram contributes to $\sigma_{\tilde{\chi}_1^0n}^{\rm SD}$ and $\sigma_{\tilde{\chi}_1^0p}^{\rm SD}$.  
Therefore, $Z$-resonant DM will be detected at LZ by SD DM-nucleon scattering, as shown in the bottom panels of Figure~\ref{fig:DD}.
Since the 90\% C.L. limit on the DM mass given by LUX~\cite{Akerib:2017kat} is about two times lower than the corresponding limit provided by PICO-60~\cite{Amole:2017dex}, while in our model $\sigma_{\tilde{\chi}_1^0n}^{\rm SD}=0.76 \sigma_{\tilde{\chi}_1^0p}^{\rm SD}$, in the following we only study the SD DM-neutron elastic cross section.

In summary, a large part of the $Z/h$ funnel region has been excluded by the current DM direct detection experimental constraints.  The surviving regions require $m_{\tilde{\chi}_1^0}\in[41,46]\cup[58,63] \, \si{\GeV}$ for positive $\mu$ and $m_{\tilde{\chi}_1^0}\in[40,46]\cup[58,63] \, \si{\GeV}$ for negative $\mu$. These regions will be probed
by the SI and SD DM-nucleon scattering detection at LZ.
We should keep in mind, however, that these regions are obtained under the assumption that the masses of all non-electroweakino sparticle masses are \SI{3}{\TeV}.  If that is not the case, for example in the case of light squarks and a light non-SM-like CP-even Higgs, the SI DM-neutron cross section could reduce and modify the allowed regions.

\subsection{\textit{Z} and Higgs boson invisible decay}
\label{sec:ZHinvisibleDecay}

If $m_{\tilde{\chi}_1^0}<M_Z/2$, the decay of $Z$ boson to a pair of neutralinos is kinematically allowed.  The decay width of this process is given by~\cite{Hamaguchi:2015rxa}:
\begin{equation}
\label{eq:decayWidthZ}
\Gamma(Z\to \tilde{\chi}_1^0\tilde{\chi}_1^0)=\frac{M_{Z}C_{Z\tilde{\chi}_1^0\tilde{\chi}_1^0}^2}{24\pi}\left(1-\frac{4m_{\tilde{\chi}_1^0}^2}{M_Z^2}\right)^{\frac{3}{2}}.
\end{equation}
$\SI{45}{\GeV}>m_{\tilde{\chi}_1^0}>\SI{40}{\GeV}$, in which DM direct detection is possible,
equation~\eqref{eq:decayWidthZ} gives $\Gamma(Z\to \tilde{\chi}_1^0\tilde{\chi}_1^0)\lesssim0.05$ MeV.  This decay width is much below the LEP bound on the new physics contribution to $\Gamma(Z\to {\rm invisible})=$ 2 MeV at 95\% C.L.  LEP bounds on electroweakino masses,
$m_{\tilde{\chi}_1^\pm}>92$ GeV and $m_{\tilde{\chi}_1^0}+m_{\tilde{\chi}_{2,3}^0}>208$ GeV~\cite{Patrignani:2016xqp}, are not constraining either in the surviving regions.

\begin{figure}[th]
\centering 
\includegraphics[width=.9\textwidth]{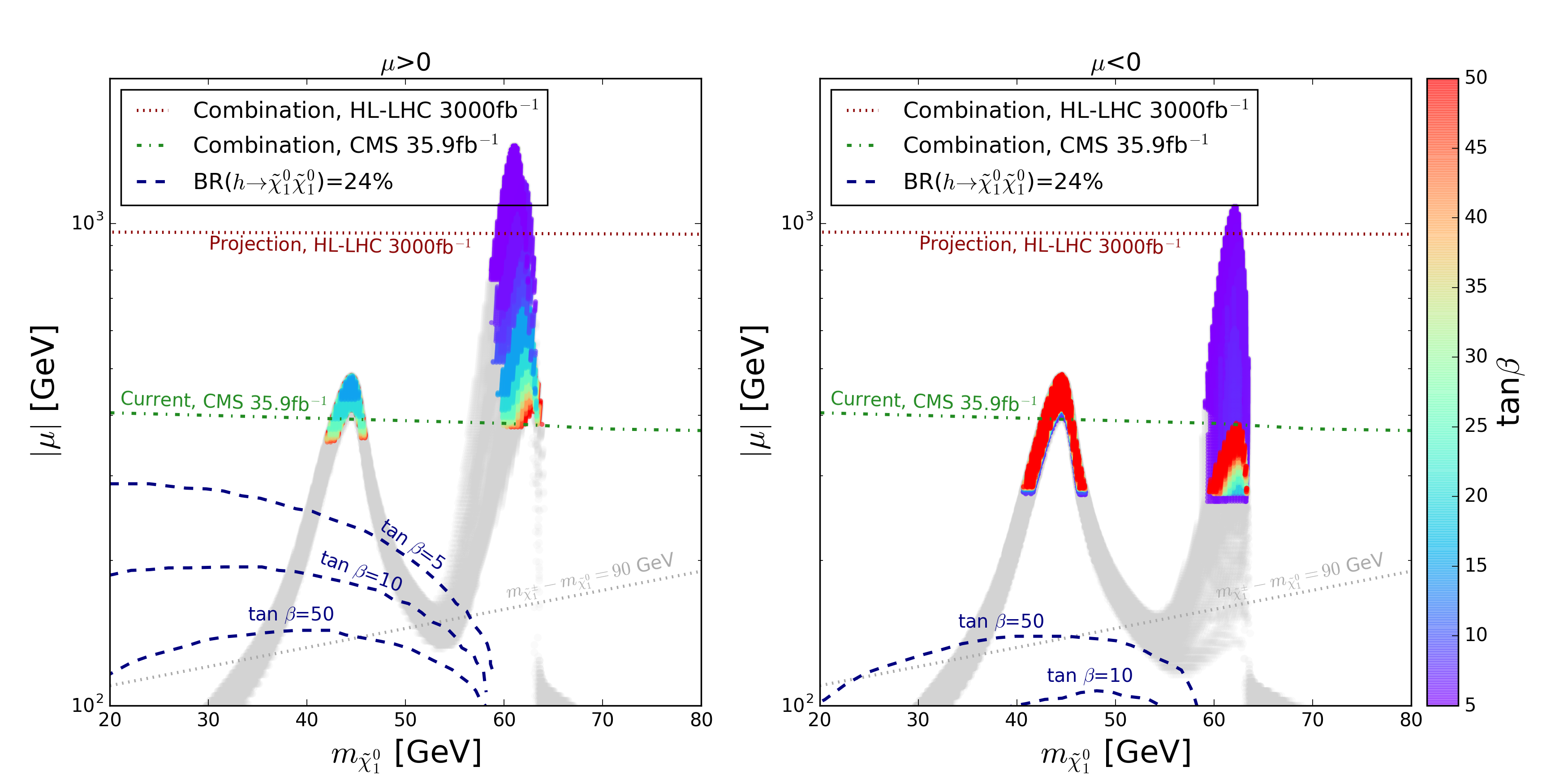}
\caption{\label{fig:13lhc} Constraints on the relevant parameter regions from invisible decay limits.  Regions excluded by DM direct detection are filled with grey colour.  The blue dashed lines indicate the 95\% C.L. upper limits on the invisible decay branching ratio of 125 GeV Higgs boson for different values of $\tan\beta$.  The green dot-dashed lines and red dotted lines show the 95\% C.L. upper limits from the combination of CMS searches for electroweakinos at the 13 TeV LHC with 35.9 fb$^{-1}$ data and at the HL-LHC with 3000 fb$^{-1}$ data, respectively.  Regions below these lines are excluded by the corresponding experimental results.}
\end{figure}

Similarly, for $m_{\tilde{\chi}_1^0}<M_h/2$, the Higgs boson decay width into a pair of neutralinos is:
\begin{equation}
\Gamma(h\to \tilde{\chi}_1^0\tilde{\chi}_1^0)=\frac{M_{h}C_{h\tilde{\chi}_1^0\tilde{\chi}_1^0}^2}{16\pi}\left(1-\frac{4m_{\tilde{\chi}_1^0}^2}{M_h^2}\right)^{\frac{3}{2}}.
\end{equation}
The combination of several searches performed by the ATLAS~\cite{Aad:2015pla} and CMS~\cite{Khachatryan:2016whc, CMS-PAS-HIG-17-023} collaborations sets an upper limit of 0.24 at the 95\% C.L. on BR$(h\to \tilde{\chi}_1^0\tilde{\chi}_1^0)$ for the 125 GeV Higgs boson.  In Figure~\ref{fig:13lhc}, we show these limits in the $(m_{\tilde{\chi}_1^0},|\mu|)$ logarithmic plane for different values of $\tan\beta$.  It is clear that the limits become stronger as $\tan\beta$ decreases (increases) for $\mu>0$ ($\mu<0$), but they are always weaker than the DM direct detection limits. The global fit of Higgs couplings will provide a stricter constraint on the invisible Higgs decay width. 
However, the constraint from global fit can be relaxed by tuning the SUSY masses that here we fix at \SI{3}{\TeV}. For instance, the best fit point of global fit for $Z/h$ funnel region in MSSM7 requires $m_{\tilde{t}_1}\simeq 2.1$ TeV and $M_A\simeq \SI{1.8}{\TeV}$~\cite{Athron:2017yua}. Thus we do not impose the Higgs invisible decay constraint from global fit in simplified model.  
The projected limit on BR$(h\to \tilde{\chi}_1^0\tilde{\chi}_1^0)$, such as BR$(h\to \tilde{\chi}_1^0\tilde{\chi}_1^0)>0.4\%$ from ILC~\cite{Asner:2013psa}, can cover the whole $Z$ funnel region, but not the $h$ funnel~\cite{Hamaguchi:2015rxa,Barman:2017swy}.

\subsection{Electroweakino searches at the 13 TeV LHC}\label{sec:13lhc}

The ATLAS~\cite{Aaboud:2018jiw, Aaboud:2018doq, Aaboud:2017leg, Aaboud:2017mpt, Aaboud:2017nhr} and CMS~\cite{Sirunyan:2017zss, Sirunyan:2017qaj, Sirunyan:2017obz, Sirunyan:2017eie,Sirunyan:2018iwl, Sirunyan:2017lae, Sirunyan:2018ubx} collaborations performed numerous searches for direct production of electroweakinos at the 13 TeV LHC.  In the simplified model in which the Wino-like $\tilde{\chi}_1^\pm$ ($\tilde{\chi}_2^0$) decays to a $W (Z)$ boson and a massless $\tilde{\chi}_1^0$, the search performed by ATLAS with 36 fb$^{-1}$ data for final states involving two or three leptons excludes Wino masses up to 580 GeV~\cite{Aaboud:2018jiw}.  The statistical combination of searches performed by CMS excludes the Wino below a mass of 650 GeV at the 95\% C.L.~\cite{Sirunyan:2018ubx}.  The corresponding mass bounds for the Higgsino might be lower than that at least 100 GeV because the production rate of Higgsino-like chargino and neutralino pair is nearly half than the production rate of Wino-like chargino and neutralino pair~\cite{Calibbi:2014lga}.  Based on these surviving regions of $Z/h$-resonance DM could be excluded since the DM relic density imposes strict requirements on the Higgsino mass, as shown in Figure~\ref{fig:13lhc}.  In the following, we assess the LHC constraints on the parameter space of interest by a detailed Monte Carlo simulation.

We use \texttt{MadGraph5\_aMC@NLO\_v2.6.1}~\cite{Alwall:2014hca} in combination with \texttt{Pythia6}~\cite{Torrielli:2010aw} to generate events for the relevant processes:
\begin{equation}
pp\to \tilde{\chi}_{1}^\pm\tilde{\chi}_{2,3}^0,~~pp\to \tilde{\chi}_{2,3}^0\tilde{\chi}_{2,3}^0,~~pp\to \tilde{\chi}_{1}^\pm\tilde{\chi}_{1}^\mp,
\end{equation}
where the production rate of the first process at the LHC is much larger than the others.  Here $\tilde{\chi}_{1}^\pm$ decays 100\% to a W boson and a $\tilde{\chi}_1^0$, $\tilde{\chi}_{2,3}^0$ decay to a $Z$ boson and a $\tilde{\chi}_1^0$ or a $h$ boson and $\tilde{\chi}_1^0$.  Although the branching ratios BR($\tilde{\chi}_{2,3}^0\to\tilde{\chi}_{1}^0Z$) and BR($\tilde{\chi}_{2,3}^0\to\tilde{\chi}_{1}^0h$) depend on $\tan\beta$ and sign of $\mu$, 
$\sum{\rm BR}(\tilde{\chi}_{2,3}^0\to\tilde{\chi}_{1}^0Z)$ and $\sum{\rm BR}(\tilde{\chi}_{2,3}^0\to\tilde{\chi}_{1}^0h)$ are roughly comparable for the whole parameter space~\cite{Calibbi:2014lga}.  
The cross sections are normalized to next-to-leading order (NLO) computed by \texttt{PROSPINO2}~\cite{Beenakker:1996ed}.  Finally, we use \texttt{CheckMATE-2.0.7}~\cite{Dercks:2016npn} with \texttt{Delphes3.4.1}~\cite{deFavereau:2013fsa} to repeat the CMS analysis~\cite{Sirunyan:2018ubx}.

The CMS combined search related to our processes~\cite{Sirunyan:2018ubx} included the following channels.
\begin{itemize}
    \item The "$\geq3 \ell$" search for the $pp\to \tilde{\chi}_{1}^\pm\tilde{\chi}_{2}^0\to W^\pm Z\tilde{\chi}_{1}^0\tilde{\chi}_{1}^0 \to \ell\ell v\ell \tilde{\chi}_{1}^0\tilde{\chi}_{1}^0$ process, with three or more leptons and large $E_T^{\rm miss}$ in the final state~\cite{Sirunyan:2017lae}.  In the several signal regions (SR) categorized by the number of lepton and lepton flavour, SR-A targets the $WZ$ topology.  This is done by selecting events with three light-flavour leptons ($e,~\mu$), two of which form an opposite-sign, same-flavour (OSSF) pair.  These events are further divided into 44 bins by the invariant mass of the pair $M_{\ell\ell}$, the transverse mass $M_T$ of the third lepton and $E_T^{\rm miss}$.  In~\cite{Sirunyan:2018ubx}, the categorization has been updated to improve the sensitivity for the region of $m_{\tilde{\chi}_2^0}-m_{\tilde{\chi}_1^0}\simeq M_Z$ by requiring $H_T$, the scalar $p_T$ sum of the jets, with $p_T>30$ GeV.  However, compared to~\cite{Sirunyan:2017lae}, the observed lower mass limit of the Wino-like $\tilde{\chi}_1^\pm$ for massless $m_{\tilde{\chi}_1^0}$ has also been improved from 450 GeV to 500 GeV.  Here we adopt the improved bins of SR-A for the analysis, but the validation of cut-flows is based on~\cite{Sirunyan:2017lae} since the cut-flow in~\cite{Sirunyan:2018ubx} has not been provided.

    \item The "$2\ell$ on-$Z$" search for the $pp\to \tilde{\chi}_{1}^\pm\tilde{\chi}_{2}^0\to ZW^{\pm} \tilde{\chi}_{1}^0\tilde{\chi}_{1}^0 \to \ell\ell jj \tilde{\chi}_{1}^0\tilde{\chi}_{1}^0$ process, with exactly two OSSF leptons consistent with the $Z$ boson mass, two non b-tagged jets consistent with the W boson mass and large $E_T^{\rm miss}$ in the final state~\cite{Sirunyan:2017qaj}.  The variable $M_{ T2}$~\cite{Lester:2014yga, Lester:1999tx} is defined using $E_T^{\rm miss}$ and the two leptons are required to be more energetic than 80 GeV to reduce the $t\bar{t}$ background.  Then four exclusive bins are defined based on $E_T^{\rm miss}$.  The analysis probes Wino-like $\tilde{\chi}_1^\pm$ masses between approximately 160 and 610 GeV for $m_{\tilde{\chi}_1^0}=0$ GeV and ${\rm BR}(\tilde{\chi}_{1}^\pm\to W^\pm\tilde{\chi}_{1}^0)={\rm BR}(\tilde{\chi}_{2}^0\to Z\tilde{\chi}_{1}^0)=100\%$.

    \item The "$1\ell2b$" search for the $pp\to \tilde{\chi}_{1}^\pm\tilde{\chi}_{2}^0\to hW^{\pm} \tilde{\chi}_{1}^0\tilde{\chi}_{1}^0 \to b\bar{b} v\ell \tilde{\chi}_{1}^0\tilde{\chi}_{1}^0$ process, with exactly one lepton, exactly two b jets and large $E_T^{\rm miss}$ in the final state~\cite{Sirunyan:2017zss}.  The invariant mass of the two b jets is required to be in the range [90, 150] GeV.  The transverse mass of the lepton-$E_T^{\rm miss}$ system and the contransverse mass $M_{\rm CT}$ of the two b jets are used to suppress backgrounds, and the $E_T^{\rm miss}$ separates the SR into two exclusive bins.  The result excludes $m_{\tilde{\chi}_1^\pm}$ between 220 GeV and 490 GeV at 95\% C.L. when the $\tilde{\chi}_1^0$ is massless in the simplified model.
\end{itemize}

Additionally, there are "H$(\gamma\gamma)$" searches for the $pp\to \tilde{\chi}_{1}^\pm\tilde{\chi}_{2}^0\to hW^{\pm} \tilde{\chi}_{1}^0\tilde{\chi}_{1}^0 \to \gamma\gamma v\ell \tilde{\chi}_{1}^0\tilde{\chi}_{1}^0$ process, and "$2\ell$ soft" searches for the $pp\to \tilde{\chi}_{1}^\pm\tilde{\chi}_{2}^0\to Z^*{W^{\pm}}^* \tilde{\chi}_{1}^0\tilde{\chi}_{1}^0 \to \ell\ell jj \tilde{\chi}_{1}^0\tilde{\chi}_{1}^0$ process where $Z^*$ and ${W^{\pm}}^*$ are off-shell.  But we do not include them in the analysis, further constraining the regions that survived DM direct detection limits, because the former can only exclude Wino below 170 GeV in a simplified model and the latter targets the situation of $m_{\tilde{\chi}_{2}^0}- m_{\tilde{\chi}_{1}^0}\simeq M_Z$.

As checked by CMS~\cite{Sirunyan:2018ubx}, these SRs are mutually exclusive, which means that they can be statistically combined to maximize the detection sensitivity.  Thus, we combine them together though the modified frequentist approach, \texttt{CL$_s$} method~\cite{0954-3899-28-10-313}, by \texttt{RooStats}~\cite{Schott:2012zb}.  The likelihood functions are written as
\begin{equation}
\mathcal{L}(\mu)= \prod_i^{N_{\rm ch}}\int d\mu' \int db_i'
\frac{(\mu' s_i+b_j')^{n_i}e^{-(\mu' s_i+b_j')}}{n_i!} \times
e^{\frac{-(\mu'-\mu)^2}{2\sigma_{\mu}^2}} \times  e^{\frac{-(b_i'-b_i)^2}{2\sigma_{b_i}^2}},
\end{equation}
where $\mu$ is the parameter of interest, $\mu'$ and $b_i'$ are nuisance parameters, and $n_i$ and $b_i$ are the number of  signal and background events in the SRs.  We take $\mu=1$ for the signal hypothesis and $\mu=0$ for the background only hypothesis.  The background event numbers $b_i$ and uncertainties $\sigma_{b_i}$ are taken from the CMS reports, while the relative uncertainties of signal $\sigma_\mu$ are assumed to equal 5\%.  Covariance matrices are not included.

\begin{table}[th]
\centering
\begin{tabular}{lcccc}
\toprule
                                                    & \texttt{BP1}        & \texttt{BP2}       & \texttt{BP3}       & \texttt{BP4}   \\
\midrule
$\tan\beta$                                          & 30                  & 10                 & 30                 & 30    \\
$M_1$ (\si{\GeV})                                              & 50             & 50             & 50              & 80     \\
$\mu$ (\si{\GeV})                                               & 390              & 390            & -390            & 390   \\
\midrule
$m_{\tilde{\chi}_1^0}$ (\si{\GeV})                             & 49.5             & 46.4           & 48.6            &  78.0 \\
$m_{\tilde{\chi}_2^0}$ (\si{\GeV})                              & 401              & 402             & 402             &  402  \\
$m_{\tilde{\chi}_3^0}$ (\si{\GeV})                              & 403              & 403             & 403             &  403  \\
$m_{\tilde{\chi}_1^\pm}$ (\si{\GeV})                            & 400              & 399             & 400             &  399 \\
BR($\tilde{\chi}_2^0\to\tilde{\chi}_1^0 Z$)         & 45\%                & 39\%               & 39\%               &  33\% \\
BR($\tilde{\chi}_2^0\to\tilde{\chi}_1^0 h$)         & 55\%                & 61\%               & 61\%               &  67\% \\
BR($\tilde{\chi}_3^0\to\tilde{\chi}_1^0 Z$)         & 63\%                & 68\%               & 69\%               &  75\% \\
BR($\tilde{\chi}_3^0\to\tilde{\chi}_1^0 h$)         & 37\%                & 32\%               & 31\%               &  35\% \\
$\sigma_{\tilde{\chi}_{2,3}^0\tilde{\chi}_{1}^\pm}$ (\si{\femto\barn}) & 59.45             & 59.48           &  59.48            &  59.46  \\
\midrule
\texttt{CL}$_s^{3l}$                                   & $0.238 \pm 0.007$   & $0.240 \pm 0.007$   & $0.251 \pm 0.007$  & $0.265 \pm 0.007$\\
\texttt{CL}$_s^{2l}$                                   & $0.266 \pm 0.018$   & $0.246 \pm 0.018$   & $0.238 \pm 0.017$  & $0.231 \pm 0.016$\\
\texttt{CL}$_s^{1l2b}$                                 & $0.549 \pm 0.009$   & $0.552 \pm 0.009$   & $0.563 \pm 0.009$  & $0.553 \pm 0.009$\\
\texttt{CL}$_s^{\rm combine}$                          & $0.049 \pm 0.005$   & $0.051 \pm 0.006$   & $0.052 \pm 0.005$  & $0.054 \pm 0.006$\\
\bottomrule
\end{tabular}
\caption{Benchmark points illustrating the result of the combined CMS electroweakino searches.  The uncertainties in \texttt{CL}$_s$ only represent the uncertainties from the \texttt{CL}$_s$ calculation and do not include the uncertainties of the signal event generation.}
\label{tab:bp}
\end{table}

In Figure~\ref{fig:13lhc} we show the 95\% C.L. combined upper limits in the plane of $m_{\tilde{\chi}_1^0}$ and $\mu$ indicated by green dot-dash lines.  They barely depend on $\tan\beta$ and the sign of $\mu$, and slightly decrease with increasing $m_{\tilde{\chi}_1^\pm}$.  To illustrate this, we choose four benchmark points of fixed $m_{\tilde{\chi}_1^\pm}$ as examples and show the details of the \texttt{CL}$_s$ in Table~\ref{tab:bp}.  Comparing \texttt{BP1}, \texttt{BP2} and \texttt{BP3} we can see that the variation of $\tan\beta$ and sign of $\mu$ will affect the branching ratios of the Higgsino-like $\tilde{\chi}_{2,3}^0$, which can be easily obtained from equation~\eqref{coupling_h}, but hardly change BR$(\tilde{\chi}_{2}^0\to\tilde{\chi}_{1}^0 Z)$+BR$(\tilde{\chi}_{3}^0\to\tilde{\chi}_{1}^0 Z)$ and BR$(\tilde{\chi}_{2}^0\to\tilde{\chi}_{1}^0 h)$+BR$(\tilde{\chi}_{3}^0\to\tilde{\chi}_{1}^0 h)$.  For \texttt{BP4}, 
a heavier Bino mass $M_1$ leads to a relatively compressed spectrum and hence smaller signal cut efficiencies.

In summary, for $Z/h$ funnel DM, regions in which $\mu$ is smaller than about 390 GeV are excluded by LHC Run-II results, which limits are stricter than DM direct detection for negative $\mu$ and positive $\mu$ with $\tan\beta>$20.  The $Z$ funnel region is on the verge of complete exclusion.  In the case of $\mu<0$, the $h$ funnel region can only survive with $\tan\beta<7.4$, while the $h$ funnel region of $\mu>0$ is the main surviving region.  The $h$ funnel regions for $\mu>0$ and $\mu<0$ are also shown in Figure~\ref{fig:hl-lhc} on the ($\tan\beta,|\mu|)$ plane to display the surviving parameter space more clearly.

\section{Electroweakino searches at the HL-LHC}
\label{sec:HL-LHC}

Although the $h$ funnel region of $\mu>0$, that is the main region that survives the current experimental limits, will be fully probed by LZ~\cite{Akerib:2018lyp}, the HL-LHC reach is still worth investigating as a complementary test. We employ two electroweakino analyses at the HL-LHC proposed by ATLAS: the "$3 \ell$" search~\cite{ATL-PHYS-PUB-2014-010} and the "$1\ell 2b$" search~\cite{ATL-PHYS-PUB-2015-032}.  Similar to the "$\geq3 \ell$" search at 13 TeV, the "$3 \ell$" search at the HL-LHC targets the $pp\to \tilde{\chi}_{1}^\pm\tilde{\chi}_{2}^0\to W^\pm Z\tilde{\chi}_{1}^0\tilde{\chi}_{1}^0 \to \ell\ell v\ell \tilde{\chi}_{1}^0\tilde{\chi}_{1}^0$ process with three or more leptons and large $E_T^\text{miss}$ in the final state.  For 3000 fb$^{-1}$ luminosity four signal regions, indicated by '\texttt{A}','\texttt{B}','\texttt{C}','\texttt{D}', optimize the  discovery and exclusion ability.  The $1\ell 2b$ search for the $pp\to \tilde{\chi}_{1}^\pm\tilde{\chi}_{2}^0\to W^\pm h \tilde{\chi}_{1}^0\tilde{\chi}_{1}^0 \to v\ell b\bar{b} \tilde{\chi}_{1}^0\tilde{\chi}_{1}^0$ process at the HL-LHC corresponds to two signal regions, '\texttt{C}' and '\texttt{D}'.
Unlike the 13 TeV analysis, the signal regions at the HL-LHC are not exclusive.  For example, in both analyses, the signal region \texttt{C} covers the signal region \texttt{D}.  As a result, we choose the signal region with the best-expected exclusion power in each analysis, and then combine them together using the \texttt{CL}$_s$ method described in Subsection~\ref{sec:13lhc}.

\begin{figure}[th]
\centering 
\includegraphics[width=.9\textwidth]{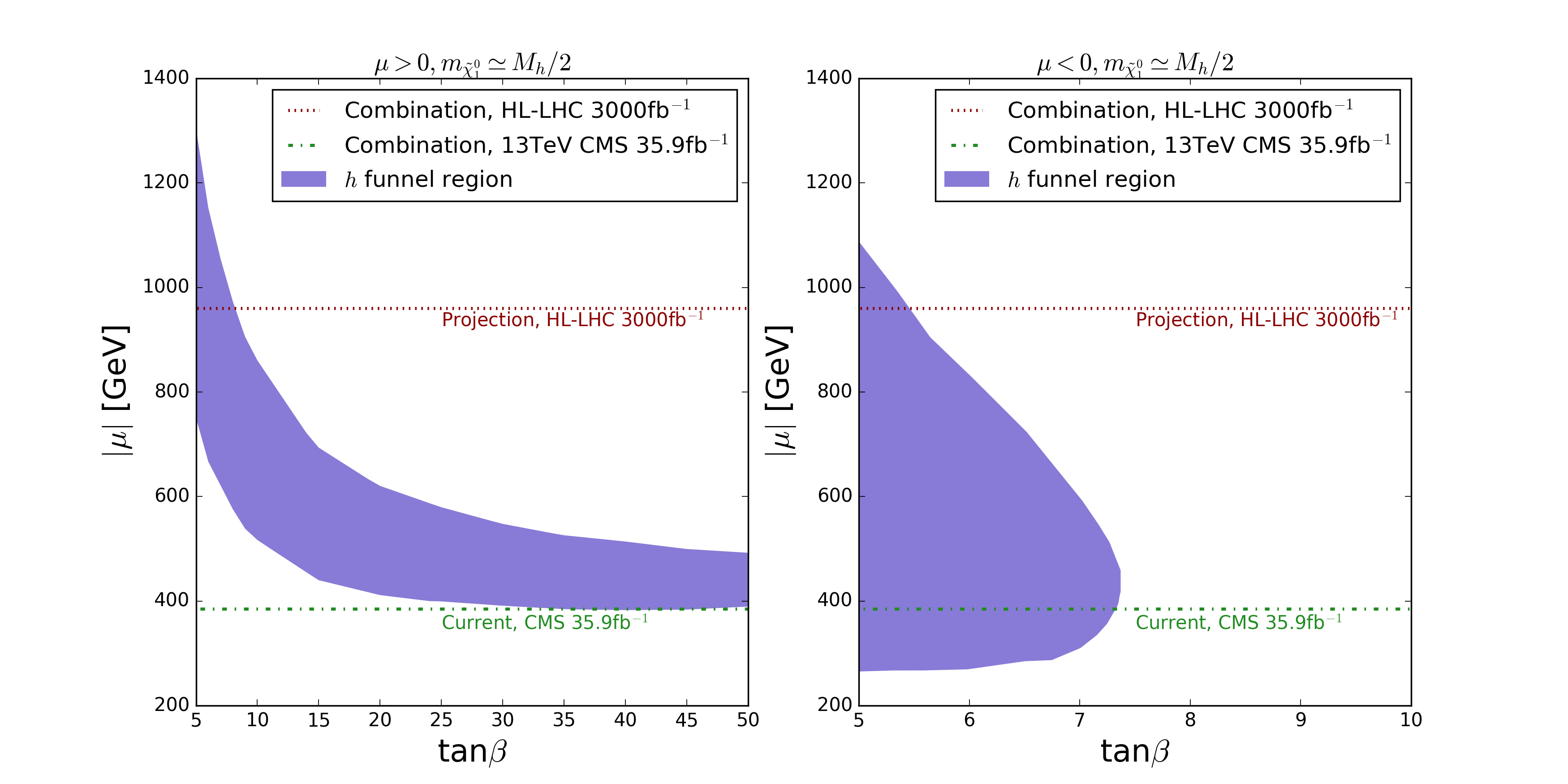}
\caption{\label{fig:hl-lhc} The plots show, in the ($\tan\beta,|\mu|$) plane, the $h$ funnel region consistent with the observed DM abundance and DM direct detection limits.  The green dot-dashed and red dotted lines show the 95\% C.L. upper limits from combined CMS searches for electroweakinos at the 13 TeV LHC with 35.9 fb$^{-1}$ data and limits by the HL-LHC with 3000 fb$^{-1}$ data, respectively.  Regions below the lines are excluded by the corresponding experimental results at 95\% C.L.}
\end{figure}

The combined expected 95\% C.L. upper limits on the $Z/h$ funnel region are presented in Figure~\ref{fig:13lhc} and Figure~\ref{fig:hl-lhc} by red dot lines.  We find that the combined result pushes the bound on $\mu$ to 960 GeV, which is 150 GeV stricter than the result of each individual analysis.  There is no doubt that the $Z$ funnel region will be completely excluded.  The parameter space of $h$ funnel region will be restricted to a very small region: $\tan\beta<8$ for $\mu>0$ and $\tan\beta<5.5$ for $\mu<0$.  Such small $\tan\beta$, however, is highly disfavoured by experimental constraints, such as the SM-like Higgs data \cite{Casas:2013pta, Arina:2014xya} and the muon anomalous magnetic moment.

\section{The \textit{Z/h} funnel in phenomenological MSSM}
\label{sec:pmssm}

After exhibiting the status of the $Z/h$
funnel region in simplified MSSM, it is desirable to 
investigate the situation when
we get rid of the assumptions, such as 
the fixed sfermion masses
and the ratio of neutralino DM to observed DM.
In this section we briefly examine the $Z/h$ funnel region in a wider model scope and with more experimental constraints. To this end, we study the light DM scenario of phenomenological MSSM (pMSSM) \cite{Cao:2015efs} by scanning the following parameter space:
\begin{equation}
\begin{gathered}
 2< \tan\beta < 60, \qquad
 10~{\rm GeV}<M_1<100~{\rm GeV}, \qquad 100~{\rm GeV} < M_2 < 1000~{\rm GeV},
 \\
 100~{\rm GeV}< \mu < 1500 ~{\rm GeV}, \qquad
 50 ~{\rm GeV} < M_A <2 ~{\rm TeV},
 \\
 |A_t=A_b| < 5 ~{\rm TeV}, \qquad
 200 ~{\rm GeV} < m_{Q_3}, \qquad
 m_{U_3}=m_{D_3} < 2 ~{\rm TeV},
 \\
 100 ~{\rm GeV} < m_{L_{1,2,3}} = m_{E_{1,2,3}} =  A_{E{1,2,3}} < 2 ~{\rm TeV}.
 \end{gathered}
 \end{equation}
The mass of the gluino and the first two generation squarks are fixed to \SI{2}{\TeV}.  In addition to the constraints described in Section~\ref{sec:constraints}, during the scan we implement the following experimental constraints at 95\% C.L.:
\begin{itemize}
    \item B-physics constraints, such as the precise measurements of $B\to X_s \gamma$, $B_s\to \mu^+\mu^-$, $B_d \to X_s \mu^+ \mu^-$ and the mass differences $\Delta M_d$ and $\Delta M_s$\cite{Patrignani:2016xqp};

    \item the muon anomalous magnetic moment ($a_{\mu}$), the measured value of which deviates from the SM prediction ($a_{\mu}^{\rm SM}$)\cite{Bennett:2006fi,Davier:2010nc};

    \item global fit of the MSSM Higgs sector implemented by the packages \texttt{HiggsBounds}~\cite{Bechtle:2013wla} and \texttt{HiggsSignals}~\cite{Bechtle:2013xfa};
    
    \item searches for direct production of charginos and neutralinos in events with $3\ell+E_T^{miss}$~\cite{Aad:2014nua} and $2\ell+E_T^{miss}$~\cite{Aad:2014vma} at LHC Run-I using \texttt{CheckMATE-2.0.23}.

\end{itemize}
We also require $m_{\tilde{l}}>2.0 \, m_{\tilde{\chi}_1^0}$ to discard the samples with DM co-annihilation through sleptons in the early universe.
Since there may be other sources of DM, here we set only an upper bound on the DM relic density.
Assuming that the other sources of the DM have no interaction with nuclei, this implies that we have to scale the DM-neutron elastic cross section by the ratio of neutralino DM relic density and observed DM abundance.

\begin{figure}[t]
\centering 
\includegraphics[width=.9\textwidth]{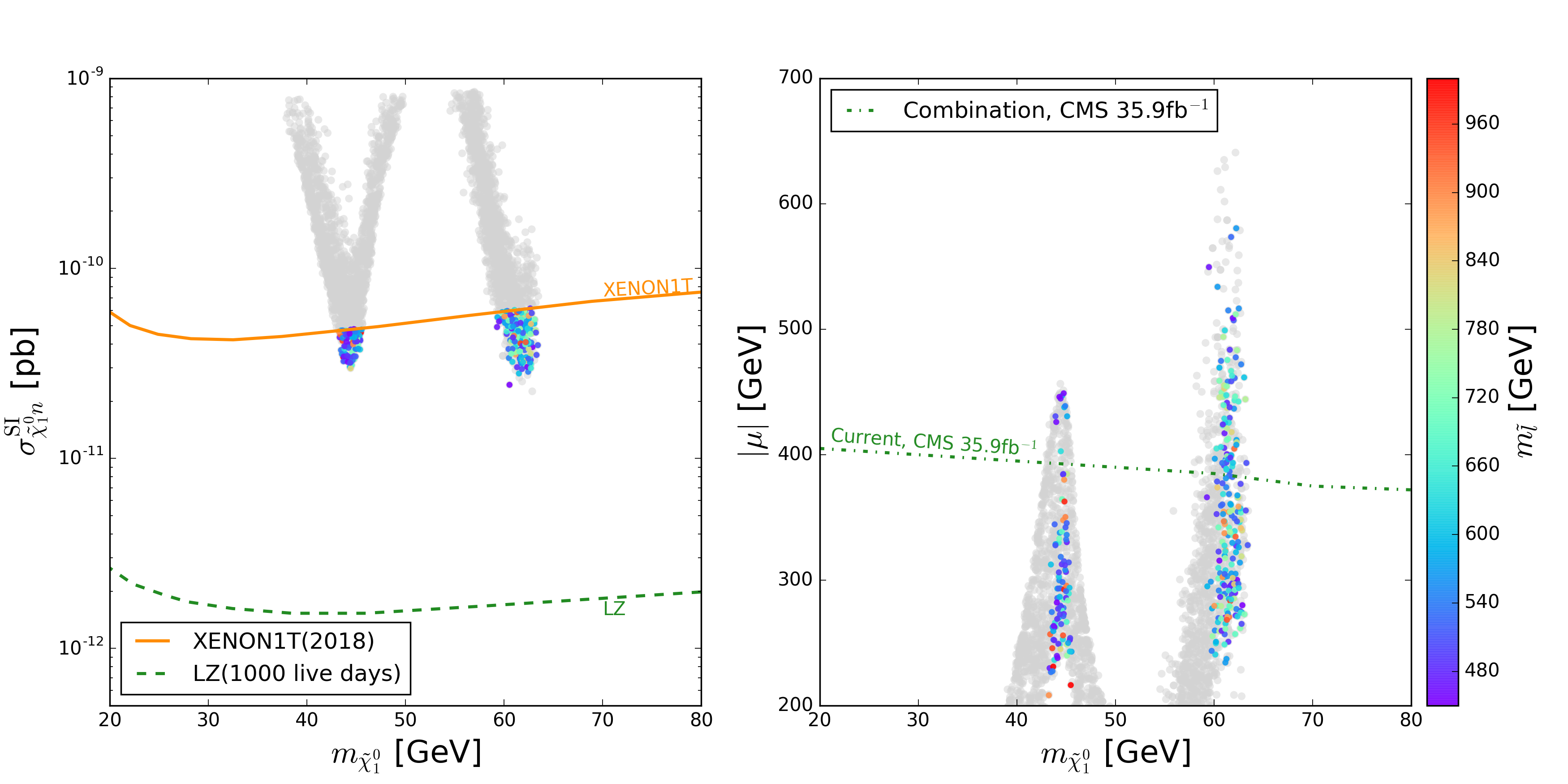}
\caption{\label{fig:pmssm} Surviving parameter regions of pMSSM shown on the lightest neutralino mass vs.\ the SI DM-neutron elastic cross section (left panel) and vs.\ the Higgsino mass (right panel).  Colours show the unified mass of sleptons, except for the grey samples that are excluded by DM direct detection at 90\% C.L. or direct searches for sleptons at LHC at 95\%C.L.}
\end{figure}

The surviving parameter regions of pMSSM are presented in Figure~\ref{fig:pmssm} with grey points indicating the samples further excluded by DM SI/SD direct detection and direct searches for sleptons using 36 fb$^{-1}$ data at LHC Run-II~\cite{Aaboud:2018jiw,Sirunyan:2018nwe}, and other colours indicating the unified mass of sleptons.  The left panel is similar to the left top panel of Figure~\ref{fig:DD}, though now $\tilde{\chi}_1^0$ may represent only part of the total DM.  Both the $Z$ and $h$ funnel regions are tightly restricted by the DM direct detection constraints that yield $m_{\tilde{\chi}_1^0}\in[43.1,45.6]$ GeV or $[59.2,63.6]$ GeV.  In the right panel we find that the combination of electroweakino searches further excludes regions where the ratio of the neutralino DM relic density over the observed DM density is smaller than 58\% (19\%) for the $Z$ ($h$) funnel region.
Comparing the pMSSM model to the simplified model we find that the constraint on $a_{\mu}$, which requires $\tan\beta>9$, reduces the height of the $h$ funnel region.
Furthermore, $a_{\mu}$ also restricts the slepton masses~\cite{Cox:2018qyi}.  As shown by the colours in Figure~\ref{fig:pmssm}, the surviving samples require either a light slepton or a light chargino.  
For $Z/h$ resonances, the points of $m_{\tilde{l}}\lesssim 460~\gev$ are excluded by the multi-lepton plus $E_T^{miss}$ searches at LHC Run-II~\cite{Aaboud:2018jiw,Sirunyan:2018nwe}, which further reduce the height of the $h$ funnel peak from 650~\gev{} to 580~\gev. Therefore, the detection of the whole $Z/h$ funnel region in pMSSM will be much faster than the one in the simplified model, in the joint result of future slepton searches and electroweakino searches at LHC. For example, if the exclusion limits on Higgsino mass and slepton mass are both improved by about 150~\gev, there would be no surviving point in pMSSM.

\section{Summary}
\label{sec:summary}
In this work we investigate the current and future status of the $Z/h$ funnel region in the MSSM with the constraints from DM direct detection, measurements of $Z/h$ invisible decay, direct searches for electroweakinos/sleptons at the LHC and muon g-2 measurement. Differently from previous studies in which the constraints from LHC were implemented by requiring the SUSY signal events in an individual signal region, we combine the results of all relevant electroweakino searches performed by the CMS, especially the "$1\ell2b$" search. Such combination increases the bound on the Higgsino mass parameter to $|\mu|>390$ GeV, which is about 80 GeV stricter than the bound obtained from individual analyses.

With such improvement, we find that in a simplified model the $Z$ funnel region is on the brink of complete exclusion, the $h$ funnel of $\mu<0$ only survives if $\tan\beta<7.4$,
and the $h$ funnel region of $\mu>0$ is the main surviving region:
\begin{enumerate}
    \item $Z$ funnel region, $m_{\tilde{\chi}_1^0}\in[42.5,45.8]$ GeV, $\mu\in[388,484]$ GeV; \label{enum:Zmu>0}
    \item $Z$ funnel region, $m_{\tilde{\chi}_1^0}\in[42.5,45.8]$ GeV, $\mu\in[-388,-486]$ GeV; \label{enum:Zmu<0}
    \item $h$ funnel region, $m_{\tilde{\chi}_1^0}\in[59.4,63.4]$ GeV, $\mu\in[-386,-1089]$ GeV, $\tan\beta\in[5,7.4]$; \label{enum:hmu<0}
    \item $h$ funnel region, $m_{\tilde{\chi}_1^0}\in[58.4,63.6]$ GeV, $\mu\in[386,1444]$ GeV. \label{enum:hmu>0}
\end{enumerate}
They can be entirely detected by LZ, while regions \ref{enum:Zmu>0} and \ref{enum:Zmu<0}, 
and most of the parameter space in region \ref{enum:hmu<0} and \ref{enum:hmu>0} can be excluded by the HL-LHC. 

In the popular pMSSM, the surviving parameter space becomes smaller due to other constraints. Especially, the light sleptons required by the muon anomalous magnetic moment will accelerate the exclusion of $Z/h$ funnel region at the LHC. Only a tiny part of the parameter space can survive the current experimental constraints. Though the modest excesses in recent electroweakino searches prefer light electroweakino, the $Z/h$ funnel region in MSSM is not
an ideal interpretation;
this is particularly true in view of a plausible improvement of the bounds on  $\sigma_{\tilde{\chi}_1^0n}^{\rm SI}$ expected by the on-going DM direct detection experiments, or also in view of the increase of the limits on slepton and electroweakino in non-compressed region by the forthcoming LHC 80 fb$^{-1}$ results.


\acknowledgments
We thank Csaba Balazs for useful comments on the draft.  This research was supported by the ARC Centre of Excellence for Particle Physics at the Tera-scale, under the grant CE110001004.

\bibliography{ldm.bib}

\end{document}